\documentclass[%
 reprint,
superscriptaddress,
 amsmath,amssymb,
 aps,
floatfix,
]{revtex4-1}

\usepackage{graphicx}
\usepackage{dcolumn}
\usepackage{bm}
\usepackage{color}

\usepackage{hyperref}

\begin{document}


\title{Emergence of Polarization in Coevolving Networks}

\author{Jiazhen Liu}
\affiliation{%
Department of Physics, University of Miami, Coral Gables, Florida 33142, USA}%
\author{Shengda Huang}
\affiliation{%
Department of Physics, University of Miami, Coral Gables, Florida 33142, USA}%
\author{Nathaniel.M.Aden}
\affiliation{%
Department of Physics, University of Miami, Coral Gables, Florida 33142, USA}%
\author{Neil.F.Johnson}
\affiliation{%
 Physics Department, George Washington University, Washington D.C. 20052, USA
}%
\author{Chaoming Song}
\email{c.song@miami.edu}
\affiliation{%
Department of Physics, University of Miami, Coral Gables, Florida 33142, USA}%




\begin{abstract}
Polarization is a ubiquitous phenomenon in social systems. Empirical studies document substantial evidence for opinion polarization across social media, showing a typical bipolarized pattern devising individuals into two groups with opposite opinions. While coevolving network models have been proposed to understand polarization, existing works cannot generate a stable bipolarized structure. Moreover, a quantitative and comprehensive theoretical framework capturing generic mechanisms governing polarization remains unaddressed. In this paper, we discover a universal scaling law for opinion distributions, characterized by a set of scaling exponents. These exponents classify social systems into bipolarized and depolarized phases. We find two generic mechanisms governing the polarization dynamics and propose a coevolving framework that counts for opinion dynamics and network evolution simultaneously. Under a few generic assumptions on social interactions, we find a stable bipolarized community structure emerges naturally from the coevolving dynamics. Our theory analytically predicts two-phase transitions across three different polarization phases in line with the empirical observations for the Facebook and blogosphere datasets. Our theory not only accounts for the empirically observed scaling laws but also allows us to predict scaling exponents quantitatively. 

\end{abstract}

\maketitle


Recently published discourse around opinion polarization, a process by which the opposition of opinions increases with time, has received much attention \cite{bramson2016disambiguation,dimaggio1996have,matakos2017measuring,manrique2018individual}. Empirical studies observe a typical bipolarized pattern where individuals are divided into two groups with radically opposite opinions. For instance, the political division between liberal and conservative parties reflects the heterogeneity in opinions regarding political orientations. These different attitudes are observed on social media \cite{Bakshy1130,cinelli2021echo,del2016spreading,garimella2018quantifying,del2016echo,borge2015content}, finding that the most frequently shared political opinions are aligned with a large proportion of the liberal or conservative population, i.e., the opinion bipolarization. On the other hand, empirical data also show a depolarization phase for some systems, where opinion distribution peaks around a neutral state with a significant variance in non-political fields \cite{Bakshy1130}.

Empirical studies show that the echo chamber effect underlies the opinion polarization in social networks \cite{cinelli2021echo,del2016spreading,garimella2018political,cota2019quantifying}, suggesting that like-minded users tend to interact \cite{Bakshy1130,cinelli2021echo,lazer2009computational}. Indeed, the coexistence of polarization of opinions and network structures implies that the interplay between opinion and network dynamics plays an essential role in polarization. While growing coevolving models are proposed to understand mutual interactions between opinion and network structures, these studies did not focus specifically on opinion polarization \cite{holme2006nonequilibrium,iniguez2009opinion,vazquez2007time,flache2011small}. 

More recently, a reinforced coevolving model has been proposed to explain network polarization \cite{baumann2020modeling,baumann2021emergence}. The reinforcement model (RM) generates either a stable mono-polarized phase where everyone leans towards one-sided opinions or a global consensus phase where everyone has the same neutral opinion. While a metastable bipolarized pattern has been observed numerically, it only appears temporarily and degenerates rapidly to a mono-polarized pattern. Moreover, its occurrence heavily relies on the initial state.

While the partial success of the RM model hints that polarization patterns might originate from the coevolving dynamics, the discrepancy between the existing models and the stable bipolarized pattern observed empirically has a deep origin. Indeed, the existing polarization modeling framework is a variation of synchronization models, where individuals synchronize to a single state when the system falls into an ordered phase. To capture a bipolarized structure, however, one requires a different framework. Moreover, coevolving network models are infamously difficult to solve due to the complicated interplay between opinion and network structure. Hence, existing models mostly rely on numerical simulations and qualitative approximations, and a quantitative model that accounts for the empirically observed bipolarized pattern is missing.   

In this Letter, we report a universal scaling law characterized by a set of scaling exponents for empirical opinion distributions. These exponents classify social systems into polarization and depolarization phases. To explain this finding, we propose two ubiquitous mechanisms for the polarization dynamics: 1) {\it Opinion homogenization.} Individuals' opinions are influenced by their neighbors in a social network and tend to converge to similar views \cite{,starnini2016emergence,sasahara2021social,clifford1973model,holley1975ergodic,black1948rationale}, and 2) {\it Homophily clustering.} Social connections evolve with time, where individuals tend to connect to those with similar beliefs. Consequently, similar individuals group together to form clusters \cite{cinelli2021echo,del2016spreading,garimella2018political,garimella2018quantifying}. These two mechanisms lead to the entanglement of network evolution with opinion dynamics. Therefore, we propose a generic coevolving framework capturing the interplay between opinion dynamics and network evolution. We find the exact solution of the proposed modeling framework and predict analytically the universal scaling laws observed in the empirical data. We calculate the phase diagrams analytically for the proposed model, predicting three stable new phases: (i) polarization, (ii) partial polarization, and (iii) depolarization, where both polarization and partial polarization phases show a stable bipolarized pattern. To our best knowledge, we offer the first coevolving network model that predicts a stable bipolarized pattern analytically, validated by numerical simulation and empirical measurements. 

\begin{figure}
  \includegraphics[width=1\linewidth]{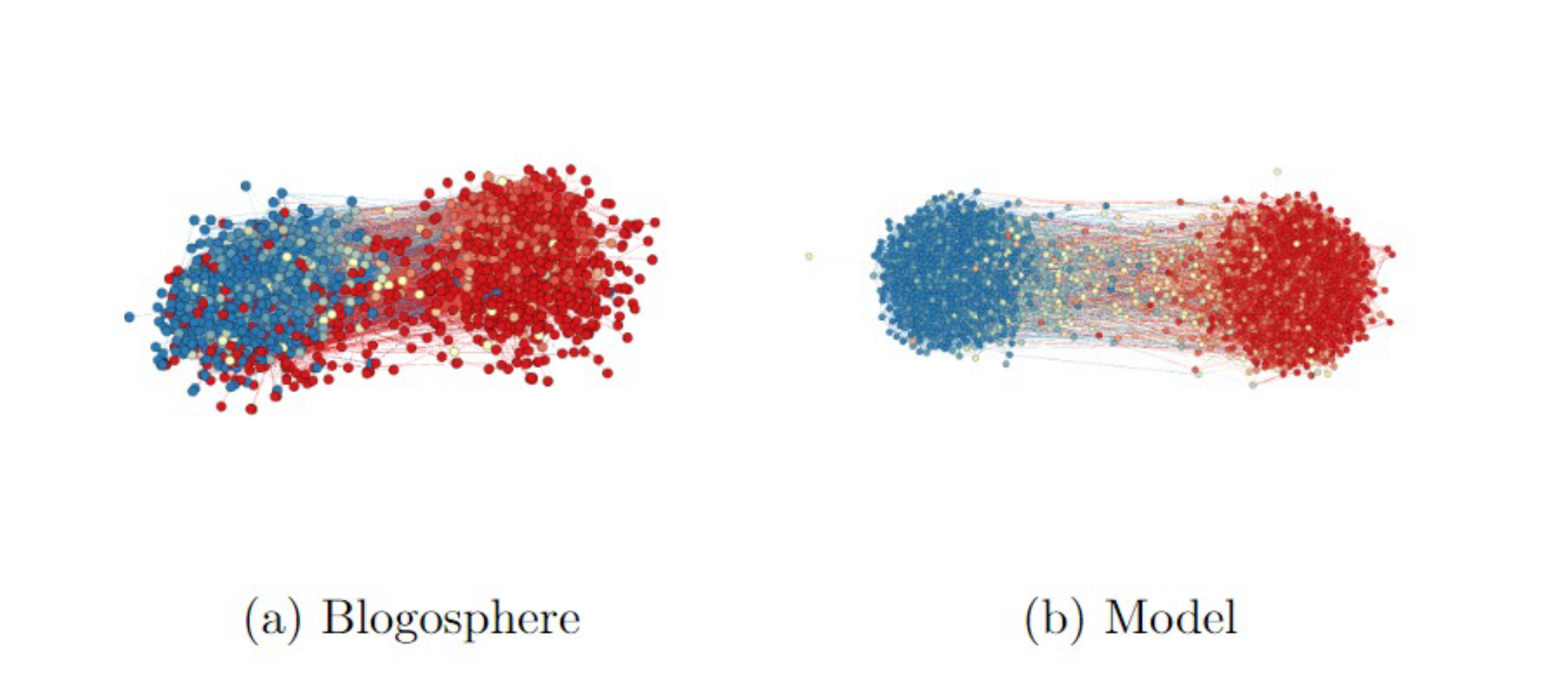}
  \caption{\textbf{Polarized networks.} (a) The Blogosphere network. Colors represent political opinions, i.e., blue for liberals and red for conservatives. (b) A social network generated by numerical simulations of our model for the polarization phase.
}
  \label{fig:cluster}
\end{figure}

{\it Experimental observations.}
We use two datasets to uncover the polarization. The first dataset consists of the 500 most shared online domains on Facebook collected \cite{Bakshy1130}. These domains are classified as either hard content (FB-HC) or soft content (FB-SC). The second dataset consists of 1,490 blogs and 19,090 references in Blogosphere \cite{adamic2005political}. We compute the mean score of political leanings $s$ for each domain or blog, where $s$ ranges from $-1$ (liberal) to $1$ (conservative) (see SM for details). Figure~\ref{fig:cluster}a depicts the Blogosphere network, suggesting that this network is polarized into two opposite communities, a phenomenon known as the echo chamber \cite{cinelli2021echo,del2016spreading,garimella2018political,cota2019quantifying,baumann2020modeling}.

\begin{figure}
\includegraphics[width=1\linewidth]{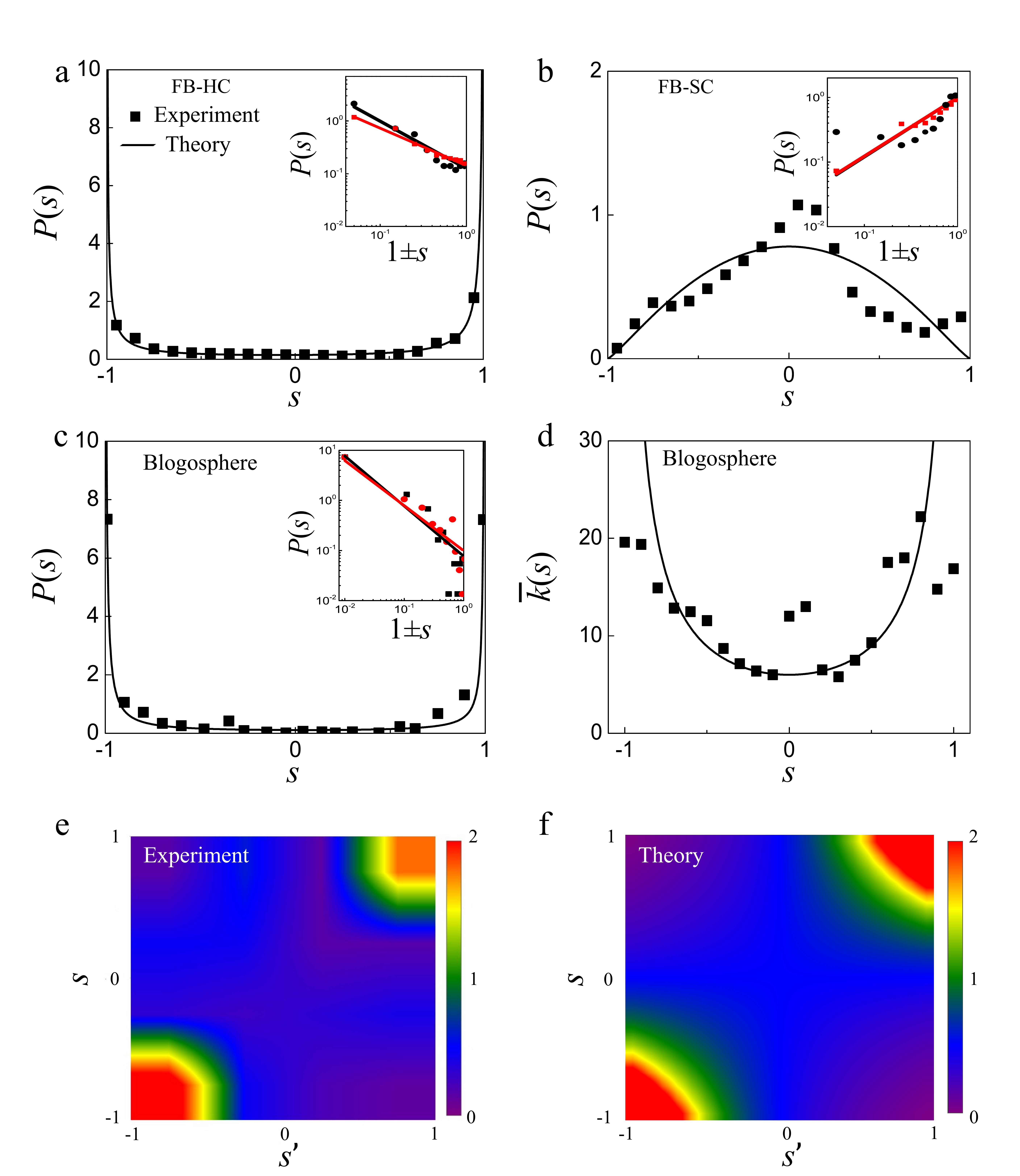}
\caption{\label{fig:model}\textbf{Empirical measures and theoretical predictions.} (a)-(c) The empirical (dots) and theoretical results (green curves) of opinion distributions for FB-HC, FB-SC, and Blogosphere. The scatter plots in the insets indicate that empirical opinion distributions follow power laws. The solid lines represent our theoretical prediction from our model. (d) The empirical results and theoretical predictions of the correlation between the network's degree and opinions for the Blogosphere. (e)-(f) Heatmaps for the empirical and theoretical normalized joint-opinion distributions.}
\end{figure}

To quantify the observed polarization, The scatter plot in Figure \ref{fig:model} depicts the opinion distribution $P(s)$ for all three empirical datasets. We find that the opinion distributions in two politics-related datasets (FB-HC and Blogosphere) are U-shaped, suggesting a polarization phase. In contrast, opinions in the FB-SC dataset are inverse U-shape distributed, indicating a depolarization phase. To investigate the scaling relation between the population size and opinion extremeness, we plot $P(s)$ as a function of $1\pm s$ in the insets of Fig.\ref{fig:model}a-c, where $1\pm s$ measures the opinion deviation from the most extreme $\pm 1$ ones. We find 
\begin{equation}
P(s) \sim \begin{cases}
(1-s)^{\delta_+} &s\to 1\\
(1+s)^{\delta_-} &s\to -1\\
\end{cases},
\label{powerlaw}
\end{equation}
satisfying power laws, where $\delta_{\pm}$ characterizes the power-law exponents when opinion $s$ approaches $\pm 1$. The negative exponent values indicate that the population increases with the extremeness of their opinions and diverges when the opinion score $s$ reaches limiting cases $\pm 1$. In contrast, positive $\delta_{\pm}$ exponents indicate $P(s)$ is peaked around $0$. Therefore, the exponents $\delta_{\pm}$ characterize polarization, i.e., $\delta_{\pm}<0$ and $>0$ for polarized and depolarized systems, respectively.

To explore the impact of the opinion's polarization on the network structure, we measured the average degree $\bar{k}$ as a function of the opinion $s$ for the Blogosphere dataset. The scatter plot in Fig. \ref{fig:model}d shows $\bar{k}$ increases with the extremeness of opinion score $s$, indicating that users who hold extreme political opinions are more likely to be the network's hubs. We further measured the joint opinion distribution $Q(s,s')$ on each edge of the Blogosphere network, which computes the number of edges connected from an individual with opinion $s$ to another with $s'$. Figure \ref{fig:model}e plots the normalized joint opinion distribution $R(s,s')=\frac{Q(s,s')}{P(s)P(s')}$, measuring the deviations from the uncorrelated distribution \cite{maslov2002specificity}. We found a strong enhancement of the number of two connecting agents with similar opinions and suppression of the number of two connecting agents with dissimilar opinions, indicating the existence of homophily clustering in the real-world social network (Fig.\ref{fig:cluster}a).\\

{\it Modeling framework.} To account for the empirical observations, we consider a coevolving network consisting of $N$ interactive agents, each with an opinion $s$, varying continuously between $-1$ and $1$. Agents' connection is described by an adjacency matrix $A$, where the matrix element $A_{ij} = 1$ or $0$ represents the agent \emph{i} being connected or disconnected to \emph{j}. Both agents' opinions and the connections among them evolve in continuous time $t$. Moreover, we require the opinion dynamics and network evolution coupled together, satisfying two intrinsic mechanisms: 1) {\it Opinion homogenization}, that is, the change of an agent's opinion is influenced by the opinions of its neighbors and tends to change to similar views. 2) {\it Homophily clustering}, that describes the social connections being more likely established if agents hold similar views. The two mechanisms are coupled dynamically with the coevolution of opinions and social connections. Below we discuss the proposed coevolving dynamics in detail. 

{\it Opinion homogenization.}
To model opinion homogenization, we assume the opinion dynamic follows
\begin{equation}\label{eq:ito}
		\mathrm{d}s_i = \mu(s_i,\vec{s},\bm A) \mathrm{d}t + \sigma (s_i) \mathrm{d}W_t,
\end{equation}
where $\vec{s}=(s_1,s_2,...s_N)$ represents the set of opinions of all agents, and $s_i$ is the opinion of agent $i$. $\bm A$ is the adjacency matrix of the agents, and $W_t$ is the standard Wiener process. $\mu$ is the drift term that controls the change of opinions on average, and $\sigma$ is the diffusion term that controls the variance of opinion dynamics. Moreover, we assume that the opinion drift $\mu$ depends on the opinions of agent $i$ and its neighbors, satisfying
\begin{equation}\label{eq:drift}
	  \mu_i=\sum_{\langle i,j\rangle}F(s_i,s_j)=\sum_{j=1}^{N}A_{ij} F(s_i,s_j),
\end{equation}
where $\langle i,j\rangle$ summing over interactions across all neighbors of $i$ \cite{holley1975ergodic,clifford1973model,deffuant2000mixing,degroot1974reaching,friedkin1990social,sood2005voter}. The pairwise force $F(s_i,s_j)$ quantifies the interaction between individuals $i$ and $j$. The opinion homogenization requires $F(s,s) = 0$ and $\partial_s F(s,s'=s) < 0$. On the other hand, the diffusion $D(s) \equiv\sigma(s)^2/2$ depends only on each agent's own opinion $s$. The boundness of opinion requires the vanishing of the diffusion at the boundaries, i.e., $D(s=\pm1)=0$, implying a Taylor expansion, $D(s)= \frac{\sigma(s)^2}{2} = \frac{\sigma_0^2}{2}(1-s^2) + O((1-s^2)^2)$. Solving Eq.(\ref{eq:ito}) leads to the time evolution for single-agent opinion
\begin{align}\label{eq:fokkerplanck}
		\frac{\partial P(s;t)}{\partial(\alpha t)}=&-\frac{\partial}{\partial s} \int_{-1}^1 F(s,s')Q(s,s';t)\,\mathrm{d}s'\notag \\
		&+\frac{\partial^2}{\partial s^2}\Big [ D(s)P(s;t)\Big ],
\end{align}
where the integral captures the neighboring interactions, since $Q(s,s';t)$ depends on the underlying network $\bm A$ implicitly.  

{\it Homophily clustering.}
To model homophily clustering, we assume that agents with similar opinions are more likely to be connected. Specifically, an agent $i$ will connect to an unlinked agent $j$ to construct a new edge with a probability rate $\gamma_+(s_i,s_j)/N$, where $s_i$ and $s_j$ are the opinions for agents $i$ and $j$, respectively. The factor $1/N$ guarantees the sparsity of the network. At the same time, an agent $i$ will disconnect with a linked agent $j$ with probability rate $\gamma_-(s_i,s_{j})$, leading to annihilating an existing edge. That is, 
\begin{subequations}
\label{eq:edge}
\begin{equation}
\mathrm{d}P[A_{ij}(0\to 1)] = \frac{\gamma_+(s_i, s_j)}{N}\mathrm{d}t\label{subeq:1}
\end{equation}
\begin{equation}
\mathrm{d}P[A_{ij}(1\to 0)] = \gamma_-(s_i, s_{j})\mathrm{d}t.\label{subeq:2}
\end{equation}
\end{subequations}
To capture the homophily clustering, we assume that $\gamma_\pm$ depends on the opinions of the agent $i$' and its neighbors $j$. This dependency leads to the coupling of network evolution with opinion dynamics. 

However, we notice that in most social media, opinion changes are much slower than network evolution. To quantify their relative time scales, we rescale the opinion drift $F\to \alpha F$ and diffusion $D\to \alpha D$ by a factor $\alpha$ that characterizes the opinion update rate relative to the network evolution. In terms of the power expansion of $\alpha$, we find the time evolution of the joint opinion distribution $Q(s,s';t)$ satisfies 
\begin{align}\label{eq:Q}
    \frac{\partial Q(s,s';t)}{\partial t}=&\gamma_+(s,s')P(s;t)P(s';t)\notag \\-&\gamma_-(s,s')Q(s,s';t)+O(\alpha),
\end{align}
where $O(\alpha)$ term captures higher-order corrections. We will keep our discussion below at the limit $\alpha \to 0$, i.e., the network evolves adiabatically, and $O(\alpha)$ will be omitted under this adiabatic approximation. 

{\it Stationary solution}. Below we will focus on the stationary solution of the proposed framework. Equation~(\ref{eq:Q}) leads to the stationary joint opinion distribution, $Q_{st}(s,s')=\frac{\gamma_+(s,s')}{\gamma_-(s,s')}P_{st}(s)P_{st}(s')$, where $P_{st}(s)$ is the stationary opinion distribution. Substituting Eq.~(\ref{eq:fokkerplanck}) leads to $\int_{-1}^{1}K(s,s')P_{st}(s')\mathrm{d}s'=\ln (D(s)P_{st}(s))$, where the kernel $K(s,s')\equiv \int \frac{\gamma_+(s,s')F(s,s')}{\gamma_-(s,s')D(s)} \mathrm{d}s$.  When $s\to\pm1$, we find $\left(\int_{-1}^{1} \kappa_\pm(s')P_{st}(s')\mathrm{d}s' \right) \ln(1\mp s) \sim \ln[(1\mp s)P_{st}(s)]$, where $\kappa_\pm(s') \equiv \frac{\gamma_+(\pm 1,s')F(\pm 1, s')}{\gamma_-(\pm 1,s')\sigma_0^2}$. Therefore, our theory predicts the universal scaling law $P_{st}(s)\sim(1\mp s)^{\delta_\pm}$, where the exponents $\delta_\pm=\int_{-1}^{1} \kappa_\pm(s')P_{st}(s')\mathrm{d}s'-1$, in line with the scaling law (\ref{powerlaw}) discovered in real-world networks (see SM Analytical Solutions). This finding suggests that the universal scaling law is rather generic for coevolving networks and largely independent of microscopic details.

{\it Opinion topology correlation.}
Equation (\ref{eq:Q}) indicates the underlying correlations between the network structures and opinions (see SM Analytical Solutions). Incorporating Eqs.~(\ref{eq:fokkerplanck}--\ref{eq:edge}) with Eq.~(\ref{eq:Q}) leads to the normalized joint opinion distribution $R(s,s') = \frac{Q(s,s')}{P(s)P(s')}=\frac{\gamma_+(s,s')}{\gamma_-(s,s')}$. Since the joint opinion distribution $Q(s,s')$ counts for the number of edges connected by two agents with opinions $s$ and $s'$. Therefore the expected degree of an agent with opinion $s$, $\bar k(s) = \int_{-1}^{1} Q(s'|s) ds' = \int_{-1}^{1} \frac{Q(s,s')}{P(s)} ds' = \int_{-1}^{1}   \mathrm{d}s'  P_{st}(s')  \frac{\gamma_+(s',s)}{\gamma_-(s',s)}$, allowing us to predict the correlation between network measure, $\bar k$, and opinion $s$.

{\it Minimal Model.}
To compare our theory with the empirical data, below we will focus on a minimal realization of the proposed framework. To be specific, we use a linear opinion dynamics model \cite{holley1975ergodic,clifford1973model}, $F(s,s') =\lambda(s'-s)$, where $\lambda$ is a constant controlling the opinion change rate. We also assume the edge birth and death rates satisfy $\gamma_{\pm}(s,s')= r_{\pm}(1+J_{\pm}ss')$, where  $|J_{\pm}|\leq 1$ quantify the strength of homophily clustering, that is, similar individuals to establish their relationships. We find the explicit form of the kernel $K(s,s') = g \int \frac{1+J_{+}ss'}{1+J_{-}ss'}\frac{2(s'-s)}{1-s^2}\mathrm{d}s $ for the minimal model, where parameter $g\equiv \frac{\lambda}{\sigma_0^2} \frac{r_+}{r_-}$. Here $\frac{\lambda}{\sigma_0^2}$ represents the opinion homogenization rate and $\frac{r_+}{r_-}$ is the birth-death ratio for network connections. The parameter $g$ integrates the interaction strengths of both opinion and network dynamics. Together with Eqs.~(\ref{eq:fokkerplanck} \& \ref{eq:Q}), we solve $P(s)$ for the minimal model ($\langle s \rangle \neq0$), finding our predictions agree with the empirical data (Fig.~\ref{fig:model}a-c, see SM Minimal Model). Note that the minimal model can also generate asymmetric $P(s)$ distribution as shown in Fig.~\ref{fig:model}a-c. However, the average opinion $\langle s \rangle$ measured on the empirical data is very close to zero, implying the corresponding $P(s)$ is well approximated by a symmetric distribution. For simplicity, we will focus only on the symmetric case ($\langle s\rangle = 0$) for the discussion below. 

\begin{figure}
  \includegraphics[width=1\linewidth]{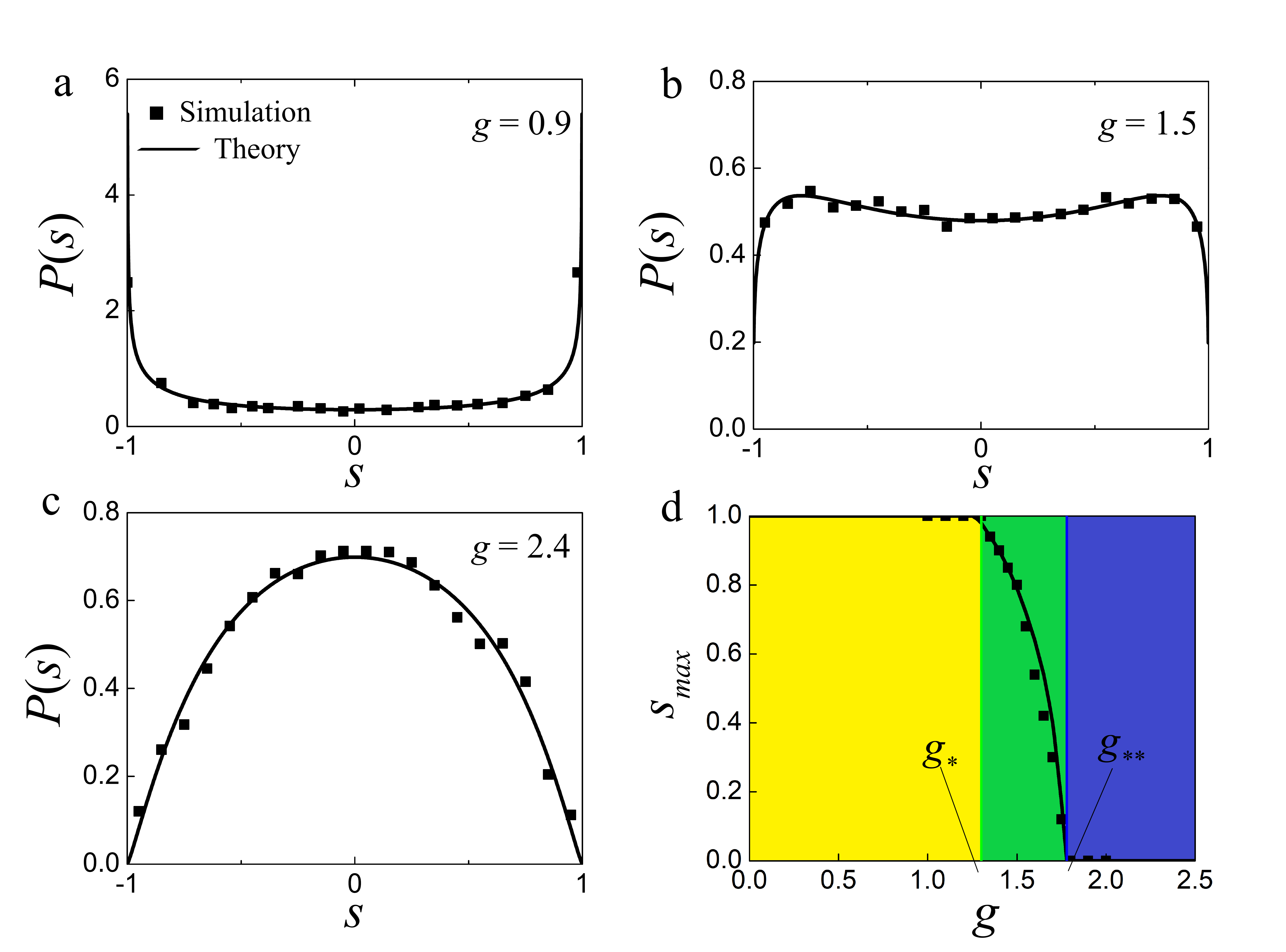}
  \caption{\textbf{Phase transitions.} (a)-(c) The phase transition of opinion distributions with fixed $J_+=0.7$ and $J_-=-0.8$. These values are determined by empirical measurements. (a) Polarization: $g=0.9$; (b) Partial polarization: $g=1.5$; and (c) Depolarization: $g=2.4$). (d) $s_{max}$ against the parameter $g$. ($g_*$: polarization (yellow) to partial polarization (green); and $g_{**}$: partial polarization to depolarization (blue)).}
  \label{fig:phasetransition}
\end{figure}

To explore different phases, we perform the numerical simulations for various $g$ with fixed $J_+$ and $J_-$. Figure 1b demonstrates a simulated minimal model of $2,000$ nodes, showing a qualitative similarity with the empirical network shown in Fig. 1a.  Figure \ref{fig:phasetransition}a shows that a bipolarized U-shaped opinion distribution $P(s)$ emerges at a smaller $g$ ($g \approx 0.9$). By gradually increasing the opinion change rate, we find that $P(s)$ transits from the U-shape to M-shape, indicating partial polarization. For sufficiently large $g$ values ($g \approx 2.4$), $P(s)$ turns into a depolarized inverse U-shaped distribution. To quantify different phases, we define an order parameter $s_{max}=|\arg\max_{s}P(s)|$ as the most probable opinion. We find $s_{max}=0$ for the depolarized phase (Fig. \ref{fig:phasetransition}c), and $s_{max}=1$ for the polarized phase (Fig. \ref{fig:phasetransition}a). For the partially polarized phase, $s_{max}$ are ranging between $0$ and $1$ (Fig. \ref{fig:phasetransition}b). Figure \ref{fig:phasetransition}d plots $s_{max}$ as a function of $g$, indicating two phase transitions exist. The first phase transition occurs at $g_*\approx1.3$, when $s_{max}$ departs from $1$, indicating that the system transits from polarization to partial polarization. The second phase transition occurs at $g_{**}\approx1.78$ when $s_{max}$ vanishes, indicating the transition between partial polarization and depolarization. Overall, decreasing $g$ moves the system towards polarization. Indeed, a smaller opinion homogenization rate drives opinions far from homogenization, heading to polarization. On the other hand, a larger birth-death ratio leads to a lower chance for dissimilar agents to be connected; hence, similar agents tend to cluster with their opinions homogenized, leading to polarization. 

To explore the opinion topology correlation, we plot predicted normalized joint opinion distribution $R(s,s') = \frac{\gamma_{+}(s,s')}{\gamma_{-}(s,s')}=\frac{r_{+}(1+J_{+}ss')}{r_{-}(1+J_{-}ss')}$ in Fig.~\ref{fig:model}f, showing two opposite clustered domains around the bottom-left and up-right corners, in line with the empirical observation (\ref{fig:model}e). Figure \ref{fig:model}d depicts the analytical prediction of $\bar k(s)$, agreeing with the empirical measurement. Note that the empirical data shows a small peak for neutral individuals, implying that these users have expected a large degree compared to our theory. This discrepancy might be due to a small group of independent media that are politically neutral but attract a large amount of connectivity from both sides. We will leave the investigation of this phenomenon for future studies.  This prediction suggests that our model generates two highly polarized clusters (Fig.\ref{fig:cluster}b), whose hubs show significant extremeness, in line with real-world polarized networks (Fig.~\ref{fig:cluster}a).   

{\it Phase diagram.}
As our theory successfully captures polarized and depolarized phases, one may wonder how the modeling parameters control different phases. We focus only on symmetric cases for simplicity, i.e., $\langle s\rangle = 0$. As we discussed above, the exponent $\delta < 0$ is for the polarized phase, whereas $\delta > 0$ is for both partial-polarized and depolarized phases. Therefore, the transition between polarization and partial-polarization emerges when the exponent $\delta = 0$, leading to ${g_*}^{-1}  + (J_+-J_-)f(J_-)  = 1$, where $f(J_-)=(1+J_-)\int^{1}_{-1}\mathrm{d}s'\frac{s'^2}{(1-J_-^2 s'^2)}P_{st}(s')$.

\begin{figure}
  \includegraphics[width=1\linewidth]{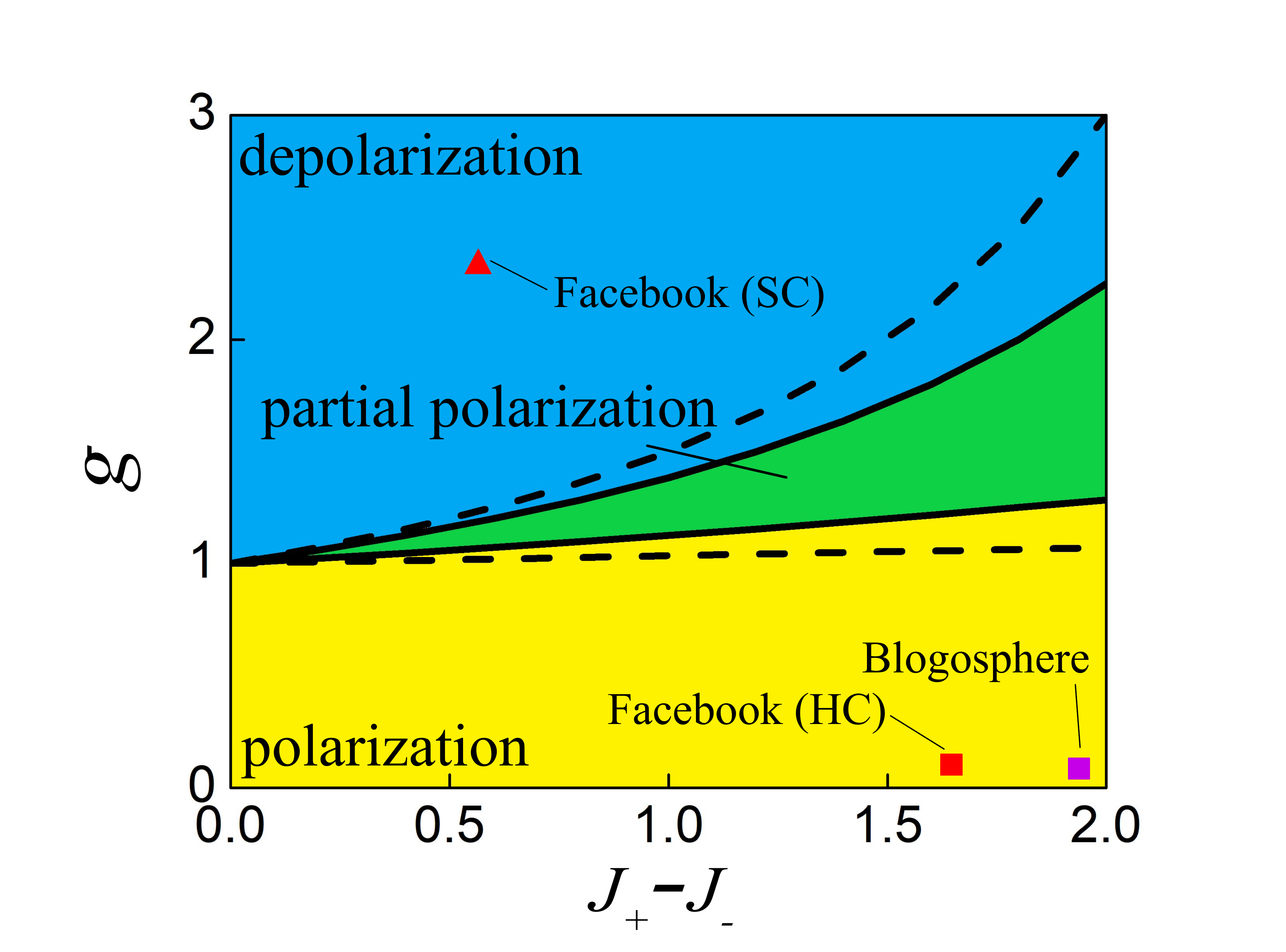}
  \caption{\textbf{Phase diagram.} This figure shows the phase transition of opinion distributions with parameters $g$ and $J_\pm$. The solid lines are the exact solutions of the phase transition lines between depolarization (blue), partial polarization (green), and polarization (yellow). The dash lines are the analytical approximations. The red and purple squares represent the real data for FB-HC and Blogosphere respectively, whereas the red triangle represents the data of FB-SC.}
  \label{fig:phasediagram}
\end{figure}

For the transition between the partially polarized and depolarized phases, the first derivative at the $s = 0$ vanishes, i.e., $P'(0) = 0$ because of the symmetry. However, for the M-shaped $P(s)$ the second derivative $P_{st}''(0)>0$, whereas $P_{st}''(0)< 0$ for inverse U-shaped $P(s)$. Therefore, the transition occurs when the second derivative vanishes, i.e., $P_{st}''(0)=0$. Substituting $P_{st}'(0)=0$ and $P_{st}''(0)=0$ into Eq.~(\ref{eq:fokkerplanck}), we obtain ${g_{**}}^{-1} + (J_+-J_-) \langle s^2\rangle = 1$, where $\langle s^2\rangle$ is the variance of the opinion distribution. Figure~\ref{fig:phasediagram} plots the phase diagram predicted by our model. The solid curves separate the domains corresponding to different phases. We mark the empirical datasets in the diagram based on the modeling parameters fitting from the data. The plot shows that Blogosphere and FB-HC are located at the polarization phase, whereas FB-SC is located at the depolarization phase.

In conclusion, we discover a universal scaling law for opinion distributions empirically, characterized by a set of scaling exponents, allowing us to quantify different polarizing phases of the real social system. We propose a generic framework for polarization dynamics of coevolving networks where opinion dynamics and network evolution are coupled based on two ingredients. Compared to the existing RM model, our model predicts stable bipolarized phases and a depolarization phase. In particular, our theory finds the bipolarized opinion distributions and network structures analytically, in line with empirical observations. Moreover, our framework offers the exact solution to coevolving network dynamics, which not only counts for the observed scaling law but also predicts the corresponding phase diagram with three different phases. 

On the other hand, our analytic solution has been found under the adiabatic approximation. We will leave the seek of a general solution for future investigations. Our theory provides a generic framework that can be applied to other areas. For instance, by introducing $s_i=\cos(\theta_i)$ the framework is capable of modeling non-linearly coupled oscillators involving background changes \cite{kuramoto1975self,sepulchre2007stabilization,wiesenfeld1998frequency}. Most importantly, our results potentially impact the understanding of human society across disciplines including social and political science.

\begin{acknowledgments}
C.S. was supported partially by the National Science Foundation under Grants 2150830 and IBSS-1620294, the Institute of Education Sciences under Grant R324A180203, and the National Institutes of Health under Grant R01DC018542. N.F.J. was funded by AFOSR grants FA9550-20-1-0382 and FA9550-20-1-0383.

J.L. and S.H. contributed equally to the work.  
\end{acknowledgments}

\providecommand{\noopsort}[1]{}\providecommand{\singleletter}[1]{#1}%

\end{document}